\pdfoutput=1
\documentclass[fleqn,usenatbib]{mnras}

\usepackage{amsmath}
\usepackage{amssymb}
\usepackage{graphicx}
\usepackage{hyperref}

\newcommand{\cadenceCompleteRatioCapOmegaDeg}{0.95}
\newcommand{\cadenceCompleteRatioCapPYr}{0.60}

\newcommand{\cadenceCompleteRatioIDeg}{0.95}
\newcommand{\cadenceCompleteRatioOmegaDeg}{0.97}
\newcommand{\cadenceCompleteRatioOmegaDot}{1.04}
\newcommand{\cadenceCompleteRatioTPeriYr}{0.48}

\newcommand{\cadenceDoubledN}{262}
\newcommand{\cadenceDoubledPctGain}{29}

\newcommand{\cadenceMcNRealizations}{30}

\newcommand{\cadenceMcOverBootOmegaDot}{0.78}

\newcommand{\cadenceReallocateNMoved}{38}
\newcommand{\cadenceReallocatePctChange}{52}
\newcommand{\cadenceReallocateRatio}{1.52}

\newcommand{\cadenceReallocateWindowEnd}{2032.75}
\newcommand{\cadenceReallocateWindowStart}{2032.16}

\newcommand{\cadenceScanCCenter}{2034.5}
\newcommand{\cadenceScanCSigmaOmegaDot}{0.00479}

\newcommand{\cadenceScanECenter}{2037.5}
\newcommand{\cadenceScanESigmaOmegaDot}{0.00239}
\newcommand{\campaignAstrometryUas}{207}

\newcommand{\campaignDenseHalfWidthDays}{20}
\newcommand{\campaignDensePct}{38.2}
\newcommand{\campaignEpochMax}{2041.69}
\newcommand{\campaignEpochMin}{2028.50}
\newcommand{\campaignNDense}{50}
\newcommand{\campaignNEpochs}{131}
\newcommand{\campaignNSparse}{81}

\newcommand{\campaignRvDeltaM}{5.3}
\newcommand{\campaignRvFluxRatio}{132}
\newcommand{\campaignRvImpliedSnr}{0.037}
\newcommand{\campaignRvPrecisionToSignal}{0.11}
\newcommand{\campaignRvResolutionElement}{60}
\newcommand{\campaignRvSigmaEris}{1621}

\newcommand{\campaignRvSignalMax}{15285}

\newcommand{\campaignSigmaDmag}{0.30}
\newcommand{\campaignSparsePct}{61.8}

\newcommand{\confusionBeamFwhmMas}{60}
\newcommand{\confusionBestFitShift}{-0.00000}
\newcommand{\confusionFloorPNineZeroHighUas}{37}
\newcommand{\confusionFloorPNineZeroLowUas}{92}

\newcommand{\confusionNMc}{20000}
\newcommand{\confusionNaiveMedianMas}{9.7}

\newcommand{\confusionSigmaRatio}{1.00}
\newcommand{\confusionSuppressionHigh}{500}
\newcommand{\confusionSuppressionLow}{200}

\newcommand{\corrnoiseCommonOneZeroZeroRatio}{2.55}

\newcommand{\corrnoiseCommonThreeZeroRatio}{1.13}

\newcommand{\corrnoiseCommonSixZeroRatio}{1.66}

\newcommand{\corrnoiseCommonModeUas}{60}
\newcommand{\corrnoiseFidBiasNsigma}{0.08}
\newcommand{\corrnoiseFidBiasOmegaDot}{-0.00027}
\newcommand{\corrnoiseFidBootOverMc}{0.71}

\newcommand{\corrnoiseFidRatioOmegaDot}{2.09}

\newcommand{\corrnoiseFidSigmaOmegaDot}{0.00349}

\newcommand{\corrnoiseNRealizations}{40}
\newcommand{\corrnoiseNRuns}{12}

\newcommand{\corrnoiseRedOneZeroZeroRatio}{2.66}

\newcommand{\corrnoiseRedFiveZeroRatio}{1.53}

\newcommand{\corrnoiseRedNoiseUas}{50}
\newcommand{\corrnoiseRedTauYr}{0.5}
\newcommand{\corrnoiseRunGapDays}{30}
\newcommand{\corrnoiseWhiteOverBootOmegaDot}{0.72}
\newcommand{\corrnoiseWhiteSigmaOmegaDot}{0.00167}

\newcommand{\budgetBottomLine}{0.0050}

\newcommand{\budgetBottomLineInflated}{0.0074}
\newcommand{\budgetBottomLineInflatedDetectionNsigma}{31}
\newcommand{\budgetBottomLineInflatedPctOfSignal}{3.20}

\newcommand{\budgetBottomLinePctOfSignal}{2.17}
\newcommand{\budgetBottomLineRatio}{2.17}
\newcommand{\budgetBottomLineWhite}{0.0024}
\newcommand{\budgetBottomLineWhiteRatio}{1.05}

\newcommand{\budgetCombinedRatio}{1.04}

\newcommand{\budgetCombinedSigma}{0.0024}

\newcommand{\budgetCorrelatedSharePct}{77}

\newcommand{\budgetInflationHigh}{1.47}
\newcommand{\budgetInflationLow}{1.22}

\newcommand{\budgetSpinNsigmaHeadlineHigh}{1.8}
\newcommand{\budgetSpinNsigmaHeadlineLow}{1.6}
\newcommand{\extmassBiasNsigmaOne}{0.06}
\newcommand{\extmassBiasNsigmaThree}{0.14}
\newcommand{\extmassBiasPctOne}{0.056}
\newcommand{\extmassBiasPctThree}{0.137}
\newcommand{\extmassBiasRateOne}{-0.00013}
\newcommand{\extmassBiasRateThree}{-0.00032}
\newcommand{\extmassMExtOneMsun}{3000}
\newcommand{\extmassMExtThreeMsun}{7500}
\newcommand{\extmassPointmassAgreementPct}{1.19}
\newcommand{\extmassRcMpc}{12.04}
\newcommand{\extmassRpMpc}{0.056}
\newcommand{\fitFitCapOmegaDeg}{73.6504}
\newcommand{\fitFitCapPYr}{8.6800}
\newcommand{\fitFitE}{0.9833}
\newcommand{\fitFitIDeg}{124.1368}
\newcommand{\fitFitOmegaDeg}{293.3197}
\newcommand{\fitFitOmegaDotDegYr}{0.2324}
\newcommand{\fitFitTPeriYr}{2023.1261}
\newcommand{\fitNBootstrap}{1000}

\newcommand{\fitPnDiffMean}{0.0005}
\newcommand{\fitPnDiffNsigma}{0.16}
\newcommand{\fitPnDiffSigma}{0.0029}
\newcommand{\fitPnFromFitMean}{0.2318}
\newcommand{\fitPnFromFitSigma}{0.0011}

\newcommand{\fitPullMax}{0.97}

\newcommand{\fitSigmaCapOmegaDeg}{0.1692}
\newcommand{\fitSigmaCapPYr}{0.0004}
\newcommand{\fitSigmaE}{0.0001}
\newcommand{\fitSigmaIDeg}{0.0728}
\newcommand{\fitSigmaOmegaDeg}{0.0953}
\newcommand{\fitSigmaOmegaDotDegYr}{0.0023}
\newcommand{\fitSigmaTPeriYr}{0.0007}
\newcommand{\fitTruthCapOmegaDeg}{73.8000}
\newcommand{\fitTruthCapPYr}{8.6800}
\newcommand{\fitTruthE}{0.9832}
\newcommand{\fitTruthIDeg}{124.0900}
\newcommand{\fitTruthOmegaDeg}{293.4000}

\newcommand{\fitTruthOmegaDotDegYr}{0.2307}
\newcommand{\fitTruthTPeriYr}{2023.1260}
\newcommand{\lensingMaxAngleBiasNsigma}{0.01}

\newcommand{\lensingMaxShiftUas}{22.2}
\newcommand{\lensingNFarSide}{4}
\newcommand{\lensingNShiftGtOneZerouas}{1}

\newcommand{\lensingOmegaDotBias}{-0.00000}
\newcommand{\lensingOmegaDotBiasNsigma}{0.00}

\newcommand{\lensingRmsShiftUas}{2.05}
\newcommand{\lensingThetaEAtMaxUas}{163}
\newcommand{\massdistDRatio}{0.9992}
\newcommand{\massdistDRatioSigma}{0.0014}
\newcommand{\massdistFitOmegaDot}{0.2320}
\newcommand{\massdistMRatio}{1.0018}
\newcommand{\massdistMRatioSigma}{0.0031}

\newcommand{\massdistOmegaDotShiftNsigma}{0.17}

\newcommand{\massdistRatioCapPYr}{0.97}
\newcommand{\massdistRatioE}{1.36}
\newcommand{\massdistRatioIDeg}{1.56}
\newcommand{\massdistRatioOmegaDeg}{1.51}
\newcommand{\massdistRatioOmegaDotDegYr}{1.04}

\newcommand{\massdistSigmaMPct}{0.97}

\newcommand{\massdistSigmaOmegaDotDegYr}{0.0024}
\newcommand{\massdistSigmaRZeroPct}{0.38}

\newcommand{\designBootRatioOmegaDotDOverA}{3.85}

\newcommand{\designCrbCapOmegaDegD}{0.1611}

\newcommand{\designCrbCapPYrD}{0.0004}

\newcommand{\designCrbED}{0.0001}

\newcommand{\designCrbIDegD}{0.0671}

\newcommand{\designCrbOmegaDegD}{0.0914}

\newcommand{\designCrbOmegaDotDegYrD}{0.00216}
\newcommand{\designCrbOverBootOmegaDotD}{0.93}

\newcommand{\designCrbTPeriYrD}{0.00072}
\newcommand{\designFloorFractionRealized}{0.382}
\newcommand{\designFloorFractionRequested}{0.40}

\newcommand{\designNBoot}{1000}

\newcommand{\designNDenseA}{118}
\newcommand{\designNDenseB}{27}
\newcommand{\designNDenseC}{28}
\newcommand{\designNDenseD}{50}

\newcommand{\designSigmaCapOmegaDegA}{0.2511}
\newcommand{\designSigmaCapOmegaDegB}{0.1351}
\newcommand{\designSigmaCapOmegaDegC}{0.1588}
\newcommand{\designSigmaCapOmegaDegD}{0.1692}
\newcommand{\designSigmaCapPYrA}{0.0003}
\newcommand{\designSigmaCapPYrB}{0.0004}
\newcommand{\designSigmaCapPYrC}{0.0005}
\newcommand{\designSigmaCapPYrD}{0.0004}
\newcommand{\designSigmaEA}{0.0001}
\newcommand{\designSigmaEB}{0.0001}
\newcommand{\designSigmaEC}{0.0001}
\newcommand{\designSigmaED}{0.0001}
\newcommand{\designSigmaIDegA}{0.0981}
\newcommand{\designSigmaIDegB}{0.0567}
\newcommand{\designSigmaIDegC}{0.0677}
\newcommand{\designSigmaIDegD}{0.0728}
\newcommand{\designSigmaOmegaDegA}{0.1924}
\newcommand{\designSigmaOmegaDegB}{0.0766}
\newcommand{\designSigmaOmegaDegC}{0.0895}
\newcommand{\designSigmaOmegaDegD}{0.0953}
\newcommand{\designSigmaOmegaDotDegYrA}{0.00890}
\newcommand{\designSigmaOmegaDotDegYrB}{0.00220}
\newcommand{\designSigmaOmegaDotDegYrC}{0.00225}
\newcommand{\designSigmaOmegaDotDegYrD}{0.00231}
\newcommand{\designSigmaTPeriYrA}{0.00054}
\newcommand{\designSigmaTPeriYrB}{0.00078}
\newcommand{\designSigmaTPeriYrC}{0.00078}
\newcommand{\designSigmaTPeriYrD}{0.00066}
\newcommand{\photcheckChiTwo}{137.5}
\newcommand{\photcheckChiTwoNsigmaFromExpected}{0.40}
\newcommand{\photcheckChiTwoReduced}{1.05}

\newcommand{\photcheckJointRatioOmegaDotDegYr}{1.03}

\newcommand{\photcheckModelAmplitudeAtEpochs}{0.112}
\newcommand{\photcheckNEpochs}{131}

\newcommand{\precisionAchievedDetectionNsigma}{96}

\newcommand{\precisionAchievedUas}{207}
\newcommand{\precisionForecastDetectionNsigma}{198}
\newcommand{\precisionForecastOmegaDot}{0.2315}
\newcommand{\precisionForecastOmegaDotSigma}{0.0012}
\newcommand{\precisionForecastSigmaCapOmegaDeg}{0.0812}

\newcommand{\precisionForecastUas}{100}

\newcommand{\refframeInjectedOffsetComponentFid}{35.4}

\newcommand{\refframeJitterOneZeroZeroRatio}{1.03}

\newcommand{\refframeJitterFiveZeroRatio}{1.03}

\newcommand{\refframeJointOffsetXFid}{33.4}

\newcommand{\refframeJointOffsetYFid}{26.6}

\newcommand{\refframeJointResidualPctFid}{0.026}

\newcommand{\refframeJointSigmaFid}{0.0024}

\newcommand{\refframeJointSigmaRatioFid}{1.04}

\newcommand{\refframeNaiveBiasCons}{-0.0010}
\newcommand{\refframeNaiveBiasFid}{-0.0005}
\newcommand{\refframeNaiveBiasNsigmaCons}{0.4}
\newcommand{\refframeNaiveBiasNsigmaFid}{0.2}
\newcommand{\refframeNaiveBiasPctCons}{0.44}
\newcommand{\refframeNaiveBiasPctFid}{0.22}
\newcommand{\refframeOffsetPriorUas}{50}
\newcommand{\refframeOffsetPriorUasConservative}{100}
\newcommand{\roemerAwareOmegaDotResidual}{1.2\times10^{-13}}
\newcommand{\roemerLtMaxDays}{5.60}
\newcommand{\roemerLtMinDays}{-0.06}
\newcommand{\roemerLtRmsDays}{3.73}
\newcommand{\roemerMaxSwingDays}{7.9}

\newcommand{\roemerNaiveCapOmegaBias}{-1.40}
\newcommand{\roemerNaiveAngleNsigmaMax}{8}
\newcommand{\roemerNaiveAngleNsigmaMin}{7}
\newcommand{\roemerNaiveIBias}{0.53}
\newcommand{\roemerNaiveOmegaBias}{-0.75}
\newcommand{\roemerNaiveOmegaDotBias}{-0.0023}
\newcommand{\roemerNaiveOmegaDotBiasNsigma}{1.0}
\newcommand{\roemerNaiveOmegaDotBiasPct}{1.00}

\newcommand{\roemerSigmaRatio}{1.00}
\newcommand{\rvtestErisPctChangeOmegaDot}{4.5}

\newcommand{\rvtestErisRvSigma}{1621}

\newcommand{\rvtestHarmoniPctChangeOmegaDot}{-4.1}

\newcommand{\rvtestHarmoniRvSigma}{20.6}

\newcommand{\rvtestHarmoniSnr}{741}

\newcommand{\lightcurveAAu}{686.7}
\newcommand{\lightcurveAAuPublished}{687.0}
\newcommand{\lightcurveAmplitudeMag}{0.115}
\newcommand{\lightcurveBandHiUm}{2.40}
\newcommand{\lightcurveBandLoUm}{1.98}
\newcommand{\lightcurveBrighteningMag}{0.075}
\newcommand{\lightcurveDimmingMag}{0.040}

\newcommand{\lightcurveRedshiftExcessPpMax}{0.72}
\newcommand{\lightcurveRedshiftExcessPpMin}{0.07}
\newcommand{\lightcurveRpRs}{136.0}
\newcommand{\lightcurveRpRsPublished}{136}

\newcommand{\lightcurveTEff}{7300}
\newcommand{\lightcurveVpKms}{25\,600}
\newcommand{\lightcurveVpPctC}{8.54}
\newcommand{\conjunctionCalendarDriftYr}{0.68}
\newcommand{\conjunctionElongBelowFourFiveEnd}{2032-02-01}
\newcommand{\conjunctionElongBelowFourFiveStart}{2031-11-06}
\newcommand{\conjunctionElongBelowEightFiveEnd}{2032-03-12}
\newcommand{\conjunctionElongBelowEightFiveStart}{2031-09-26}
\newcommand{\conjunctionElongMinDate}{2031-12-19}
\newcommand{\conjunctionElongMinDeg}{5.6}
\newcommand{\conjunctionGapDaysAfterPeri}{131}
\newcommand{\conjunctionGapDaysBeforePeri}{22}
\newcommand{\conjunctionGapTotalDays}{153}
\newcommand{\conjunctionPeriOneDate}{2031-10-22}
\newcommand{\conjunctionPeriOneElongDeg}{59}
\newcommand{\conjunctionPeriTwoElongDeg}{170}

\newcommand{\spinLtNsigmaNoiseChiOneZeroZero}{5.7}

\newcommand{\spinLtNsigmaNoiseChiNineZero}{5.2}
\newcommand{\spinLtNsigmaRefChiOneZeroZero}{5.5}

\newcommand{\spinLtNsigmaRefChiNineZero}{5.0}
\newcommand{\spinLtPctChiOneZeroZero}{5.74}

\newcommand{\spinLtPctChiNineZero}{5.17}
\newcommand{\spinLtRateChiOneZeroZero}{0.01325}

\newcommand{\spinLtRateChiNineZero}{0.01192}

\title[Catching Up with S301]{Catching Up with the Fastest Star in the Galaxy: A Two-Passage GRAVITY+ Campaign to Detect S301's Schwarzschild Precession}

\author[J. Catanzarite]{J. Catanzarite$^{1,2}$\thanks{E-mail: jcatanz3@jhu.edu}
\\
$^{1}$Whiting School of Engineering, Johns Hopkins University, 3400 North Charles Street, Baltimore, MD 21218, USA\\
$^{2}$Nova Scholar Education
}

\pubyear{2026}

\begin{document}
\label{firstpage}
\pagerange{\pageref{firstpage}--\pageref{lastpage}}
\maketitle

\begin{abstract}
S301 is the fastest known star in the Galaxy. It orbits Sgr~A$^*$ every 8.68~yr on an $e = 0.9832$ orbit that carries it within 136 Schwarzschild radii at \lightcurveVpKms{}~km~s$^{-1}$ \citep{gravity2026}. An orbit like that must precess, by the same Schwarzschild effect that turns Mercury's perihelion, whatever the spin of the black hole. We ask whether an astrometry-only GRAVITY+ campaign covering the 2031.8 and 2040.5 periapsis passages can catch it. We validate a relativistic light-curve model, design a \campaignNEpochs{}-epoch Fisher-optimal cadence with a periapsis floor, fit a seven-parameter orbit including the light-travel-time delay, and build a systematic error budget. With white noise at the 207~$\mu$as achieved on S301, the fit recovers the injected rate, \fitTruthOmegaDotDegYr{}~deg~yr$^{-1}$, as $\fitFitOmegaDotDegYr \pm \fitSigmaOmegaDotDegYr$~deg~yr$^{-1}$ ($\pm\precisionForecastOmegaDotSigma$ at the forecast 100~$\mu$as). Real astrometric noise is not white. We add per-run calibration offsets and a wandering Sgr~A$^*$ photocentre at GRAVITY's published amplitudes, marginalize over the reference zero point, mass and distance, add the bounded systematics in quadrature, and inflate by the excess scatter of real GRAVITY orbit fits. What survives is \budgetBottomLineInflated{}~deg~yr$^{-1}$, a \budgetBottomLineInflatedDetectionNsigma{}$\sigma$ detection. The Sun hides the first periapsis for \conjunctionGapTotalDays{}~d, at a cost to the period and periapsis epoch but not to the precession rate. Radial velocities are out of reach at S301's magnitude; photometry only validates the light-curve model. The spin of Sgr~A$^*$ could shift the rate by \spinLtPctChiNineZero{}--\spinLtPctChiOneZeroZero{}~per~cent and cannot be bounded, so the campaign is a spin-agnostic test of general relativity, and a forecast until the star comes round.
\end{abstract}


\section{Introduction}

The S-stars around Sgr~A$^*$ have delivered the strongest tests of general relativity in the field of a supermassive black hole: the gravitational redshift and the Schwarzschild precession of S2 \citep{gravity2018,gravity2020}. S301 \citep{gravity2026} is a sharper probe: half the period of S2 (8.68 against $\sim$16~yr), a higher eccentricity, and a periapsis ten times closer (136 against $\sim$1400 Schwarzschild radii, $R_S$). It was found in a search for a star that could measure the spin of Sgr~A$^*$ through Lense--Thirring precession; a companion paper \citep{piran2026} develops the multi-star method that measurement requires.

This study began with a simpler question. Picture the star at closest approach, 136~$R_S$ from the horizon and moving at 8~per~cent of the speed of light. Some of the light it sends our way must pass so near the black hole that it is bent aside or swallowed. Would we see the star dim as it grazes the shadow? Section~\ref{sec:lightcurve} answers: the black hole takes almost nothing, and the star does change brightness, but through beaming and gravitational redshift, and by too little to detect at nineteenth magnitude. The relativistic signal worth chasing is in the astrometry. Any Schwarzschild black hole, spinning or not, turns the long axis of an orbit around it; for S301 the predicted rate is \fitTruthOmegaDotDegYr{}~deg~yr$^{-1}$, an order of magnitude faster than for S2. Can a campaign that could be proposed today catch it? Three things stand in the way. S301 is faint ($m_K = 19.3$), so only interferometric astrometry can follow it and no current spectrograph can measure its velocity. The field is crowded. And its next periapsis, in October 2031, falls behind the Sun.

We answer by building the complete pipeline an observing proposal would need: the adopted orbit (Section~\ref{sec:solution}); a validated light-curve model, the choice of instrument and observables, the cadence and the orbit fit (Section~\ref{sec:pipeline}); the recovered precession and tests of the cadence (Section~\ref{sec:results}); the systematic error budget, where the feasibility case is decided (Section~\ref{sec:budget}); and what the test does and does not claim (Section~\ref{sec:discussion}). Every number below is produced by the accompanying code and inserted by macro, so the prose cannot drift from the calculation.

\section{The S301 orbit}
\label{sec:solution}

All ground-truth parameters come from \citet{gravity2026}, Extended Data Table~2, Solution~1 (Table~\ref{tab:params}); a second, nearly degenerate solution is published but not used. The discovery fit already includes the R{\o}mer (light-travel-time) delay and Schwarzschild precession, so these elements are defined in the same relativistic model we use. At periapsis S301 passes 136~$R_S$ from Sgr~A$^*$ at \lightcurveVpKms{}~km~s$^{-1}$, more than 8~per~cent of $c$, closer and faster than any other known star. For comparison, S2 reaches $\approx$7650~km~s$^{-1}$ at its own periapsis \citep{gravity2018}; the fastest star previously claimed near Sgr~A$^*$, S4714, reaches $23\,900 \pm 8\,800$~km~s$^{-1}$ \citep{peissker2020}; and the fastest hypervelocity star known, S5-HVS1, is leaving the Galaxy at $1755 \pm 50$~km~s$^{-1}$ \citep{koposov2020}. S301 is the fastest known star in the Galaxy. Its effective temperature is unpublished; we adopt \lightcurveTEff{}~K from standard F1.5V tables, which affects only the light-curve amplitude. The mass and distance of Sgr~A$^*$ are those of \citet{gravity2022massdist}: $M_\bullet = (4.297\pm0.012)\times10^6\,M_\odot$ and $R_0 = 8277\pm9$~pc (statistical), with systematic uncertainties of $\approx40{,}000\,M_\odot$ and $\approx30$~pc.

\begin{table}
\caption{Adopted S301 parameters \citep{gravity2026}, Solution 1.}
\label{tab:params}
\begin{tabular}{lcc}
\hline
Parameter & Value & Uncertainty \\
\hline
Period $P$                 & 8.680 yr        & 0.110 yr \\
Eccentricity $e$           & 0.9832          & 0.0010 \\
Inclination $i$            & $124.09^\circ$  & $1.10^\circ$ \\
Node $\Omega$              & $73.8^\circ$    & $3.5^\circ$ \\
Argument of periapsis $\omega$ & $293.4^\circ$ & $2.2^\circ$ \\
Periapsis epoch $t_0$      & 2023.126 yr     & 0.010 yr \\
Semi-major axis $a$        & 83.0 mas (687 AU) & -- \\
Sgr~A$^*$ mass $M_\bullet$ & $4.297\times10^6\,M_\odot$ & -- \\
Distance $R_0$             & 8277 pc         & -- \\
$K$ magnitude $m_K$        & 19.3            & 0.3 \\
\hline
\end{tabular}
\end{table}

\begin{figure}
\centering
\includegraphics[width=0.7\columnwidth]{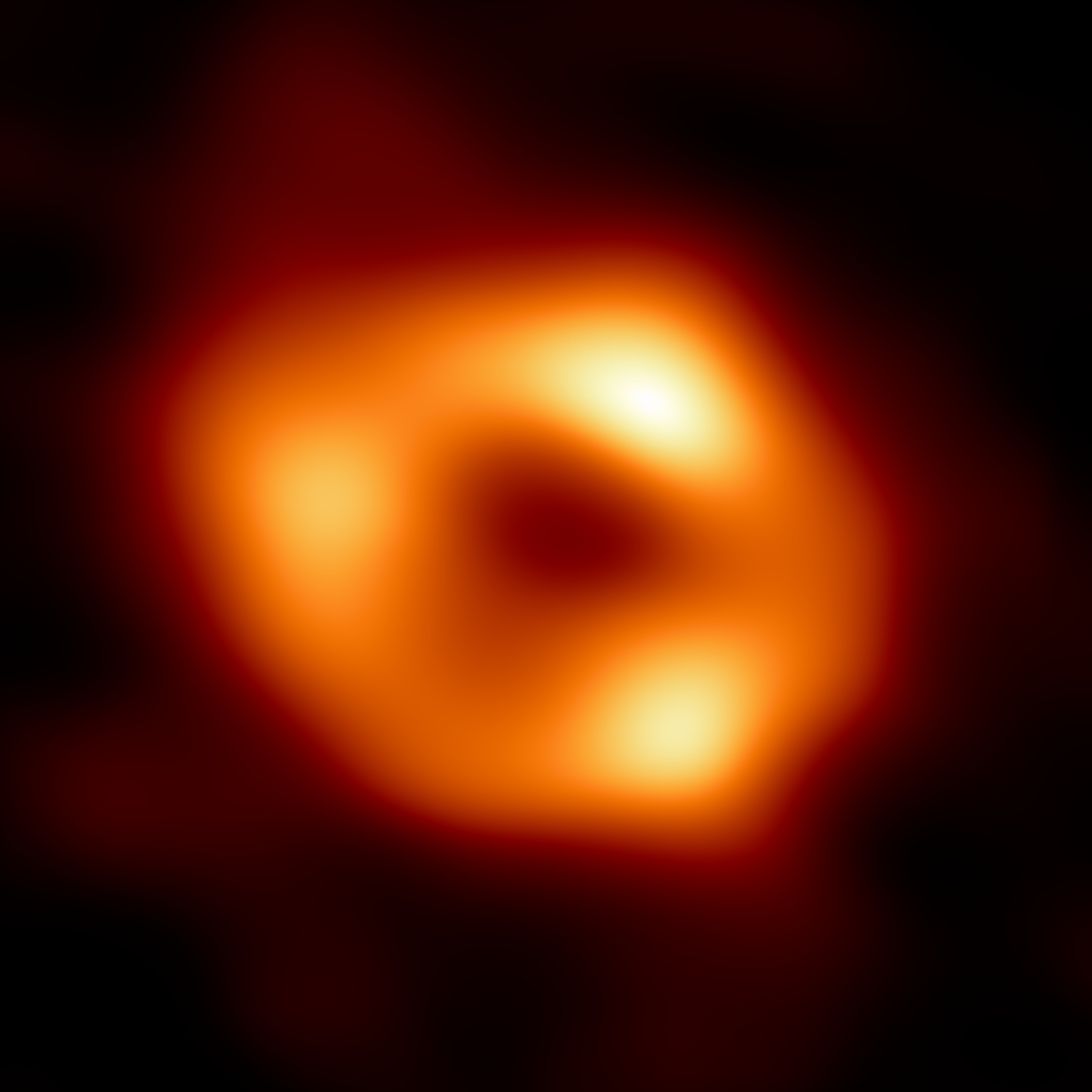}
\caption{Sgr~A$^*$ itself, for scale: the Event Horizon Telescope image of the black hole's shadow, ring diameter $51.8\pm2.3$~$\mu$as \citep{eht2022}, about 1600 times smaller than S301's 83~mas orbit and far below the astrometric resolution of this campaign. Image credit: EHT Collaboration, released by the European Southern Observatory (ESO) as eso2208-eht-mwa (2022-05-12), licensed CC~BY~4.0.}
\label{fig:ehtshadow}
\end{figure}

\begin{figure}
\centering
\includegraphics[width=\columnwidth]{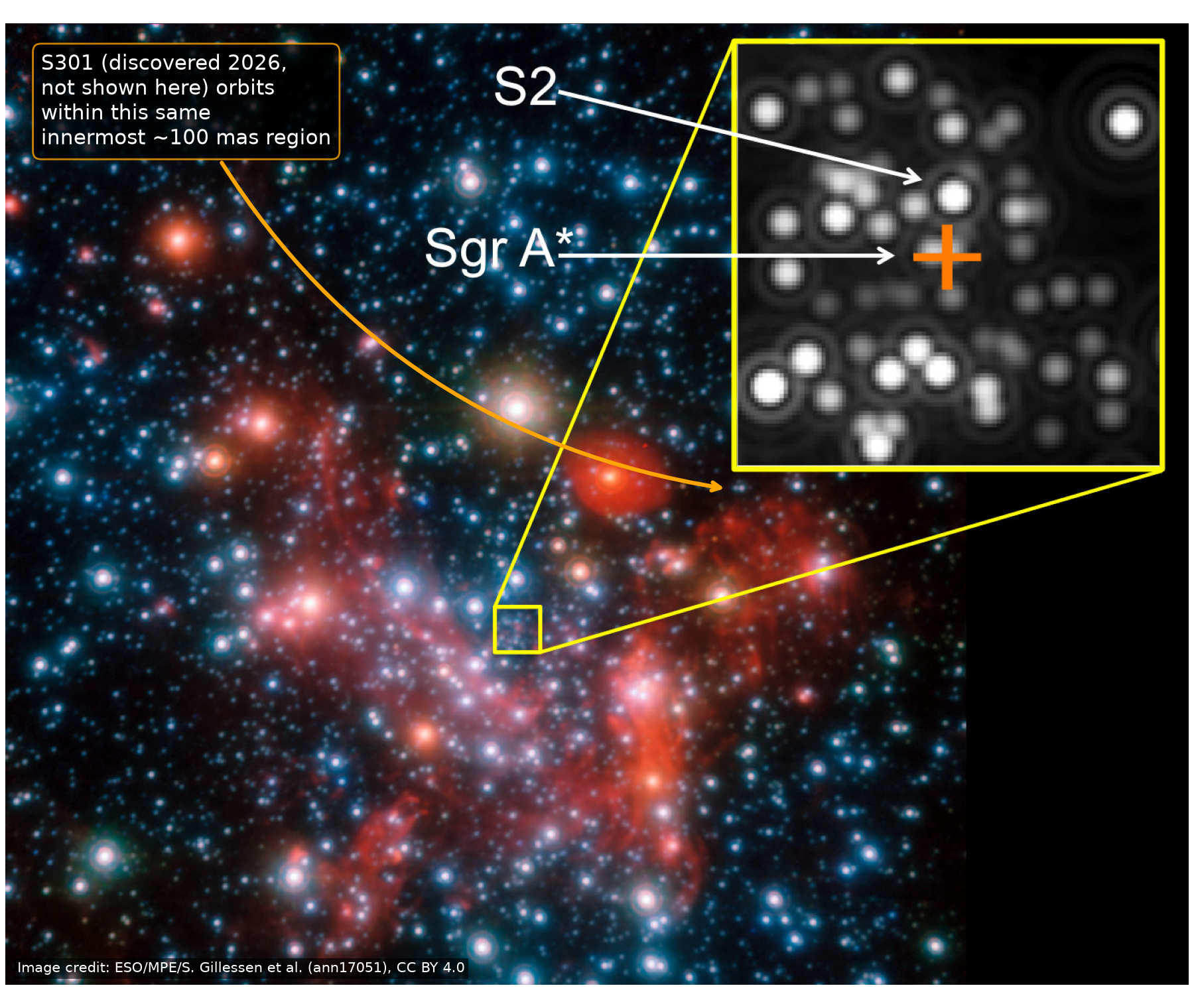}
\caption{The crowded innermost Galactic Centre in the near-infrared (a real image), with the S2/Sgr~A$^*$ zoom inset from the original release. S301 was discovered in 2026, after this 2017 image, and is not resolved in it; the schematic annotation shows that its 83~mas orbit lies in the same innermost few tens of mas, the crowding relevant to Section~\ref{sec:confusion}. Image credit: ESO/MPE/S.\ Gillessen et al., released as ann17051 (2017-08-09), licensed CC~BY~4.0.}
\label{fig:context}
\end{figure}

\section{Method}
\label{sec:pipeline}

\subsection{Relativistic light curve}
\label{sec:lightcurve}

We model S301's $K$-band light curve from the orbit of Table~\ref{tab:params}. The model applies the exact special-relativistic Doppler factor and the Schwarzschild gravitational redshift $\sqrt{1-R_S/r}$ to a \lightcurveTEff{}~K blackbody, boosts the intensity as $I_\nu/\nu^3$, and integrates over a top-hat bandpass covering GRAVITY's $K$-band range, \lightcurveBandLoUm{}--\lightcurveBandHiUm{}~$\mu$m \citep{gravity2017}. Three effects are omitted on purpose: lensing by Sgr~A$^*$, treated as an astrometric systematic in Section~\ref{sec:lensing}; spin-dependent (Kerr) terms, which would require a value for the unmeasured spin; and the R{\o}mer delay, which does not change a curve plotted against time from periapsis and enters the fit instead (Section~\ref{sec:fit}). The model reproduces the published solution: $a = \lightcurveAAu{}$~AU against \lightcurveAAuPublished{}, periapsis distance \lightcurveRpRs{}~$R_S$ against \lightcurveRpRsPublished{}, and periapsis speed \lightcurveVpPctC{}~per~cent of $c$ against the published ``$>$8\%''. This validates the engine before we ask it to predict anything new.

\begin{figure*}
\centering
\includegraphics[width=0.72\textwidth]{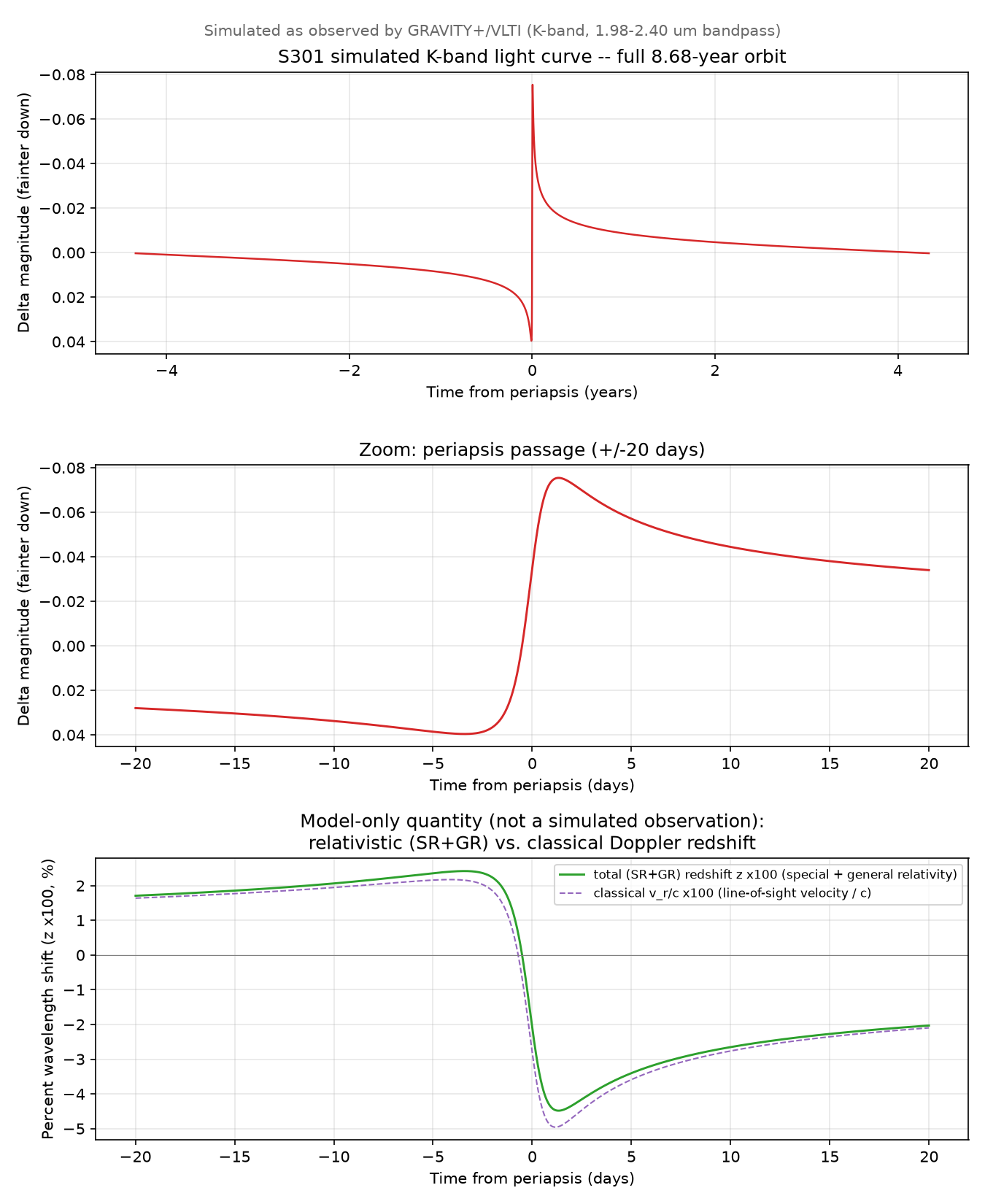}
\caption{Simulated relativistic $K$-band light curve of S301 over one orbit (top), the periapsis passage (middle), and the relativistic and classical spectral shifts near periapsis (bottom). The lower panels show a generic passage in time from periapsis; precession changes only the orbit's orientation on the sky, so the shape is the same for either real passage. The bottom panel is a model-internal quantity (SR: special-relativistic Doppler shift; GR: gravitational redshift) that explains the magnitude curves above it; no instrument in this campaign measures a spectral shift.}
\label{fig:lightcurve}
\end{figure*}

The predicted periapsis signature is an asymmetric beaming spike (Fig.~\ref{fig:lightcurve}). The star brightens by \lightcurveBrighteningMag{}~mag on approach and fades to \lightcurveDimmingMag{}~mag below baseline on recession, when de-boosting and gravitational redshift act together; the peak-to-trough amplitude is \lightcurveAmplitudeMag{}~mag. The spectral slope in the band sets the amplitude: at 2.2~$\mu$m and \lightcurveTEff{}~K, $h\nu/kT\approx0.9$, so the in-band flux scales roughly as $(1+z)^{-1.5}$, far more gently than the bolometric $(1+z)^{-4}$. Near periapsis the full relativistic redshift exceeds the classical Doppler shift by \lightcurveRedshiftExcessPpMin{}--\lightcurveRedshiftExcessPpMax{} percentage points.

The dimming deserves care, because it is not the dimming we first imagined. Sgr~A$^*$ neither swallows nor deflects a measurable share of S301's light: at 136~$R_S$ the horizon subtends a negligible fraction of the star's sky, and Section~\ref{sec:lensing} shows that light bending amounts to an astrometric shift of at most \lensingMaxShiftUas{}~$\mu$as at a handful of epochs. The fading after periapsis is gravity acting on the light in a subtler way. Each photon loses energy climbing out of the potential well, on top of the Doppler de-boosting of a receding source; the black hole taxes the photons it lets through rather than taking them. There is no occultation to model, because Sgr~A$^*$ is compact and blocks nothing.

\subsection{Instrument and observables}
\label{sec:campaign}

GRAVITY+ is the upgraded beam combiner of the VLTI (Very Large Telescope Interferometer) that discovered S301. It combines the four 8.2~m Unit Telescopes at Paranal, each with adaptive optics, on baselines up to $\sim$130~m; the baseline, not the aperture, gives it sub-milliarcsecond astrometry in the crowded Galactic Centre. We adopt a per-epoch precision of \campaignAstrometryUas{}~$\mu$as, the value achieved on S301 itself in the discovery data, where it is set by the pixel grid of the image reconstruction \citep[Methods]{gravity2026}. The discovery paper's own forecast for the completed GRAVITY+ is 100~$\mu$as, ``as expected for the final performance of GRAVITY+'' \citep{gravity2026}; we treat that as the optimistic case and report it alongside (Section~\ref{sec:results}). Each epoch also yields $K$-band photometry, which we simulate with S301's published uncertainty of \campaignSigmaDmag{}~mag but do not fit (Section~\ref{sec:photometry}).

Radial velocity was rejected before any data were simulated. SINFONI, the integral-field spectrograph formerly at the VLT (Very Large Telescope), reached 12.3~km~s$^{-1}$ on S2 at $m_K=14.0$ \citep{gravity2018}; its successor ERIS (Enhanced Resolution Imager and Spectrograph) carries the SPIFFIER spectrograph at resolving power $R=5000$ \citep{davies2023}. S301 is \campaignRvDeltaM{}~mag fainter than S2 ($\sim$\campaignRvFluxRatio{}$\times$ less flux), so in the background-limited regime the SINFONI benchmark scales to $\sim$\campaignRvSigmaEris{}~km~s$^{-1}$ per epoch: \campaignRvPrecisionToSignal{} of S301's $\pm$\campaignRvSignalMax{}~km~s$^{-1}$ velocity swing, and only \campaignRvImpliedSnr{} of SPIFFIER's \campaignRvResolutionElement{}~km~s$^{-1}$ resolution element. A line centroid cannot be located to so small a fraction of a resolution element at this flux, and the discovery team found the same at the telescope: ``Our current ERIS spectroscopy is not deep enough'' \citep{gravity2026}. Section~\ref{sec:rv} confirms by simulation that an ERIS channel would not improve any parameter and assesses the ELT (Extremely Large Telescope) prospects.

\subsection{Cadence design}
\label{sec:cadence}

An orbit this eccentric is mostly waiting. S301 drifts through apoapsis for the better part of a decade and then sweeps through most of its angular track in a few weeks around periapsis; a cadence that misses those weeks has watched the star stand still. Where the epochs fall in time therefore decides what the fit can recover. We compared four designs at a fixed budget of \campaignNEpochs{} epochs, all restricted to Sgr~A$^*$'s observing season from Paranal ($\sim$March--September; \citealt{gravity2026}).

\emph{Design A (heuristic)} places about 90~per~cent of the epochs within $\pm$\campaignDenseHalfWidthDays{}~d of each periapsis and spreads the rest thinly.

\emph{Design B (D-optimal)} lets the statistics choose. The Fisher information predicts how much each candidate epoch shrinks the joint uncertainty on the seven fitted parameters; a D-optimal design minimizes the volume of the joint uncertainty ellipsoid, evaluated at one assumed-true parameter vector and built greedily from a daily, season-filtered candidate grid.

\emph{Design C (pseudo-Bayesian D-optimal)} averages the Fisher information over 300 draws from the published element uncertainties \citep{chaloner1995}, discounting epochs that are informative only if the orbit is already known exactly.

\emph{Design D (constrained)} runs Design C's search after reserving \designFloorFractionRequested{} of the budget for the $\pm$\campaignDenseHalfWidthDays{}~d windows around both periapses. The season filter removes roughly half of passage~1's window, so \designFloorFractionRealized{} of the budget survives there.

Table~\ref{tab:cadence} gives each design's bootstrap precision, the analytic Cram\'er--Rao bound (CRB) at the truth, and the number of epochs near periapsis. Design~A is the outlier: its $\dot\omega$ precision is \designBootRatioOmegaDotDOverA{}$\times$ worse than Design~D's, because a periapsis-heavy budget leaves too little angular diversity to fix the orbit's orientation. Designs~B, C and~D agree on $\dot\omega$ within bootstrap noise, and all place many epochs near periapsis unprompted (\designNDenseB{}, \designNDenseC{} and \designNDenseD{} of \campaignNEpochs{}). We adopt Design~D because its floor guarantees periapsis coverage whatever the search does, and complete periapsis coverage is what pins the period and periapsis epoch (Section~\ref{sec:cadencetests}). For Design~D the CRB and the bootstrap agree (CRB/bootstrap = \designCrbOverBootOmegaDotD{} for $\dot\omega$), the best available evidence that the bootstrap uncertainties are sound.

The floor has a history. An earlier version of the code, with no season filter and an unsafeguarded Kepler solver, produced a Design~B that piled epochs at the ends of the campaign, skipped periapsis and failed its own bootstrap. That pitfall of an unfiltered candidate grid on a nonlinear periodic problem does not recur in the current code; we keep the floor as insurance. \citet{piran2026}'s multi-star design faces a different constraint, synchronizing observations across several stars to subtract Newtonian confusion; with one star and one observable, periapsis coverage is what matters.

\begin{table*}
\caption{Cadence designs at $N=\campaignNEpochs{}$: bootstrap $1\sigma$ uncertainties (\designNBoot{} resamples) for Designs A--D, the Cram\'er--Rao bound for the adopted Design D, and the published uncertainties. Epochs within $\pm$\campaignDenseHalfWidthDays{}~d of a periapsis: A \designNDenseA{}, B \designNDenseB{}, C \designNDenseC{}, D \designNDenseD{}.}
\label{tab:cadence}
\begin{tabular}{lcccccc}
\hline
Parameter & A & B & C & D & D (CRB) & Published \\
\hline
$P$ (yr)                 & \designSigmaCapPYrA{} & \designSigmaCapPYrB{} & \designSigmaCapPYrC{} & \designSigmaCapPYrD{} & \designCrbCapPYrD{} & 0.1100 \\
$e$                      & \designSigmaEA{} & \designSigmaEB{} & \designSigmaEC{} & \designSigmaED{} & \designCrbED{} & 0.0010 \\
$i$ (deg)                & \designSigmaIDegA{} & \designSigmaIDegB{} & \designSigmaIDegC{} & \designSigmaIDegD{} & \designCrbIDegD{} & 1.1000 \\
$\Omega$ (deg)           & \designSigmaCapOmegaDegA{} & \designSigmaCapOmegaDegB{} & \designSigmaCapOmegaDegC{} & \designSigmaCapOmegaDegD{} & \designCrbCapOmegaDegD{} & 3.5000 \\
$\omega$ (deg)           & \designSigmaOmegaDegA{} & \designSigmaOmegaDegB{} & \designSigmaOmegaDegC{} & \designSigmaOmegaDegD{} & \designCrbOmegaDegD{} & 2.2000 \\
$t_0$ (yr)               & \designSigmaTPeriYrA{} & \designSigmaTPeriYrB{} & \designSigmaTPeriYrC{} & \designSigmaTPeriYrD{} & \designCrbTPeriYrD{} & 0.0100 \\
$\dot\omega$ (deg yr$^{-1}$) & \designSigmaOmegaDotDegYrA{} & \designSigmaOmegaDotDegYrB{} & \designSigmaOmegaDotDegYrC{} & \designSigmaOmegaDotDegYrD{} & \designCrbOmegaDotDegYrD{} & -- \\
\hline
\end{tabular}
\end{table*}

\subsection{Solar conjunction}
\label{sec:conjunction}

The season filter exposes a hard limit. Sgr~A$^*$'s solar elongation is below $45^\circ$, and the field unobservable from any facility, from \conjunctionElongBelowFourFiveStart{} to \conjunctionElongBelowFourFiveEnd{}, reaching \conjunctionElongMinDeg{}$^\circ$ on \conjunctionElongMinDate{}. The first periapsis (\conjunctionPeriOneDate{}, elongation \conjunctionPeriOneElongDeg{}$^\circ$) falls inside this window. The last observable epoch before it is \conjunctionGapDaysBeforePeri{}~d before periapsis and the first after it is \conjunctionGapDaysAfterPeri{}~d after: a \conjunctionGapTotalDays{}-d blackout covering the periapsis and the start of the recovery. No facility escapes it. The elongation stays below $85^\circ$, the sunshield limit of JWST (James Webb Space Telescope), from \conjunctionElongBelowEightFiveStart{} to \conjunctionElongBelowEightFiveEnd{}; the ELT site is 20~km from Paranal and shares its sky; and the limit is geometric, so a spectrograph sees no better than an imager. The second periapsis, at elongation \conjunctionPeriTwoElongDeg{}$^\circ$, is comfortably in season. The Sun, not the black hole, decides which passage we see. With $P = 8.68$~yr each periapsis arrives \conjunctionCalendarDriftYr{}~yr later in the calendar than the last, and in 2031 the calendar puts it behind the Sun; in 2040 it does not. Section~\ref{sec:cadencetests} quantifies the cost.

\subsection{Orbit fit}
\label{sec:fit}

We fit seven parameters to the astrometry: the six Keplerian elements and the apsidal precession rate $\dot\omega$, applied as $\omega_{\rm eff}(t) = \omega + \dot\omega\,(t - t_{\rm ref})$. The model evaluates the orbit at each epoch's emission time, solved from the arrival time by Newton iteration, so the R{\o}mer delay is in both the truth and the model, as in the discovery analysis. The fit is Levenberg--Marquardt least squares started from the published solution, perturbed by 3~per~cent in period, 0.03 in eccentricity, $10^\circ$ in inclination, $15^\circ$ in node, $12^\circ$ in argument of periapsis, 0.05~yr in periapsis epoch and 0.1~deg~yr$^{-1}$ in precession rate, to stand in for the uncertainty of a preliminary orbit. The semi-major axis follows from $(P, GM_\bullet)$ by Kepler's third law with $M_\bullet$ and $R_0$ fixed at their published values; Section~\ref{sec:massdist} tests the cost of freeing them. Uncertainties come from \fitNBootstrap{} bootstrap resamples of the epochs, cross-checked against the CRB (Table~\ref{tab:cadence}) and against \cadenceMcNRealizations{} independent noise realizations (Section~\ref{sec:cadencetests}). We compare the recovered $\dot\omega$ with the Schwarzschild prediction,
\begin{equation}
\dot\omega = \frac{6\pi\,GM_\bullet}{c^2\,a\,(1-e^2)\,P},
\label{eq:precession}
\end{equation}
the Mercury-perihelion formula applied to S301. We inject no Lense--Thirring term, so $i$ and $\Omega$ stay fixed: the spin of Sgr~A$^*$ is unmeasured, and assuming a value would substitute a guess for a measurement.

\section{Results}
\label{sec:results}

\subsection{The recovered precession}
\label{sec:recovered}

\begin{figure*}
\centering
\includegraphics[width=0.72\textwidth]{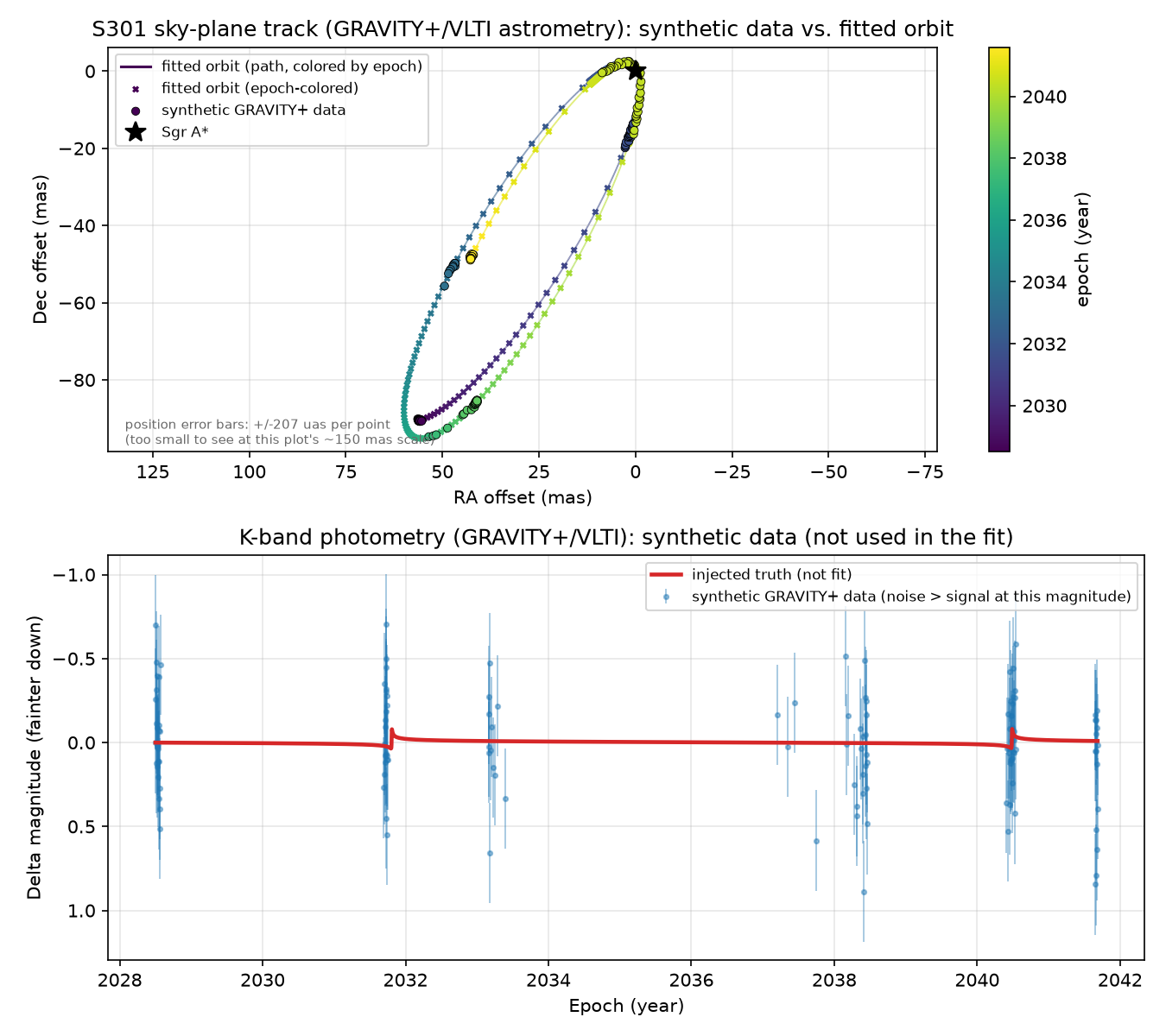}
\caption{Top: fitted sky-plane track against the synthetic GRAVITY+ astrometry, coloured by epoch, showing the two passages of the precessing orbit. Bottom: simulated $K$-band photometry against the injected truth curve (not used in the fit). An animated version (\texttt{s301\_precession.mp4}, an ancillary file of this submission, also at \url{https://github.com/jcatanza/s301-schwarzschild-precession/blob/master/output/s301_precession.mp4}) extrapolates the same precession rate over 20 orbits ($\sim$174~yr) to make the rosette visible; the two orbits analysed here are highlighted and the extrapolation is labelled on screen throughout.}
\label{fig:fitorbit}
\end{figure*}

The adopted campaign has \campaignNEpochs{} epochs between \campaignEpochMin{} and \campaignEpochMax{}. Of these, \campaignNDense{} (\campaignDensePct{}~per~cent) fall within $\pm$\campaignDenseHalfWidthDays{}~d of a periapsis, above Design~D's realized floor, because the Fisher search placed some of its flexible budget near periapsis too. The remaining \campaignNSparse{} epochs (\campaignSparsePct{}~per~cent) form a few tight clusters separated by multi-year gaps (Fig.~\ref{fig:fitorbit}), in 2033 and 2037--2038 rather than immediately after passage~1; Section~\ref{sec:cadencetests} tests whether that is wasteful.

Table~\ref{tab:finalfit} gives the fit. Every parameter is recovered within \fitPullMax{}$\sigma$ of the truth, and the fitted sky-plane track (Fig.~\ref{fig:fitorbit}, top) shows the periapsis direction rotated between passages, the geometric signature of precession. Several uncertainties are far tighter than the published ones ($\fitSigmaCapOmegaDeg^\circ$ against $3.5^\circ$ on $\Omega$, for example) because of three idealizations: the same precision at every epoch, two full passages the 2026 data set does not yet have, and fixed $M_\bullet$ and $R_0$, which pin the orbit's angular scale (Section~\ref{sec:massdist}). The comparison is not a claim to outperform the discovery data.

If GRAVITY+ reaches its forecast precision (\precisionForecastUas{}~$\mu$as instead of \precisionAchievedUas{}~$\mu$as), the same campaign gives $\dot\omega = \precisionForecastOmegaDot \pm \precisionForecastOmegaDotSigma$~deg~yr$^{-1}$, a \precisionForecastDetectionNsigma{}$\sigma$ detection instead of \precisionAchievedDetectionNsigma{}$\sigma$, with the orientation angles improved in proportion ($\sigma_\Omega = \precisionForecastSigmaCapOmegaDeg^\circ$). Precision scales linearly with the per-epoch error, so Table~\ref{tab:finalfit} can be rescaled to whatever GRAVITY+ delivers in 2031. These are white-noise figures; Section~\ref{sec:budget} is where they meet reality.

The fitted $\dot\omega$ is a free parameter, determined by geometry with no reference to equation~(\ref{eq:precession}). As a check, for each bootstrap resample we evaluate that formula on the resample's own fitted $(P, e)$, still blind to any published or injected value: the prediction is $\fitPnFromFitMean \pm \fitPnFromFitSigma$~deg~yr$^{-1}$, and its per-resample difference from the fitted $\dot\omega$ is $\fitPnDiffMean \pm \fitPnDiffSigma$~deg~yr$^{-1}$, \fitPnDiffNsigma{}$\sigma$ from zero (the difference is taken per resample because $P$, $e$ and $\dot\omega$ come from the same correlated resamples). Geometry and theory agree without either referring to the other's answer.

\begin{table}
\caption{Final result: Design D cadence, astrometry only, \campaignAstrometryUas{}~$\mu$as per epoch, \fitNBootstrap{} bootstrap resamples.}
\label{tab:finalfit}
\setlength{\tabcolsep}{4pt}
\begin{tabular}{lcccc}
\hline
Parameter & Truth & Fitted & $\sigma$ & Published $\sigma$ \\
\hline
$P$ (yr)                 & \fitTruthCapPYr{}    & \fitFitCapPYr{}    & \fitSigmaCapPYr{} & 0.1100 \\
$e$                      & \fitTruthE{}    & \fitFitE{}    & \fitSigmaE{} & 0.0010 \\
$i$ (deg)                & \fitTruthIDeg{}  & \fitFitIDeg{}  & \fitSigmaIDeg{} & 1.1000 \\
$\Omega$ (deg)           & \fitTruthCapOmegaDeg{}   & \fitFitCapOmegaDeg{}   & \fitSigmaCapOmegaDeg{} & 3.5000 \\
$\omega$ (deg)           & \fitTruthOmegaDeg{}  & \fitFitOmegaDeg{}  & \fitSigmaOmegaDeg{} & 2.2000 \\
$t_0$ (yr)               & \fitTruthTPeriYr{} & \fitFitTPeriYr{} & \fitSigmaTPeriYr{} & 0.0100 \\
$\dot\omega$ (deg yr$^{-1}$) & \fitTruthOmegaDotDegYr{} & \fitFitOmegaDotDegYr{}   & \fitSigmaOmegaDotDegYr{} & -- \\
\hline
\end{tabular}
\end{table}

\subsection{Testing the cadence}
\label{sec:cadencetests}

Two features of the adopted cadence invite suspicion, the truncated first passage and the late clusters, and we tested both. First, does the conjunction gap cost anything? We built a design identical to Design~D except that passage~1's cluster is a complete, symmetric $\pm$\campaignDenseHalfWidthDays{}~d window centred on periapsis, and bootstrapped both. Completing the passage tightens $P$ and $t_0$ to \cadenceCompleteRatioCapPYr{}$\times$ and \cadenceCompleteRatioTPeriYr{}$\times$ their Design~D values and the orientation angles by a few per cent (\cadenceCompleteRatioIDeg{}, \cadenceCompleteRatioCapOmegaDeg{} and \cadenceCompleteRatioOmegaDeg{}$\times$), while $\dot\omega$ stays at \cadenceCompleteRatioOmegaDot{}$\times$, unchanged within bootstrap noise. The gap costs the period and periapsis epoch and nothing measurable on $\dot\omega$, whose precision comes from comparing the orbit's orientation across the $\sim$8.7~yr between the two periapsis-anchored clusters.

Second, given that the gap is unavoidable, where should the freed budget go? Moving the \cadenceReallocateNMoved{} flexible epochs into the earliest window the season permits (\cadenceReallocateWindowStart{}--\cadenceReallocateWindowEnd{}, starting \conjunctionGapDaysAfterPeri{}~d after periapsis) makes $\dot\omega$'s precision \cadenceReallocateRatio{}$\times$ Design~D's, \cadenceReallocatePctChange{}~per~cent worse, and every placement tried between 2032 and 2037 followed the same pattern (best at \cadenceScanECenter{}, \cadenceScanESigmaOmegaDot{}~deg~yr$^{-1}$; worst at \cadenceScanCCenter{}, \cadenceScanCSigmaOmegaDot{}). Watching S301 for another month after periapsis means watching it slow down along a path the dense cluster has already traced. The fit learns more from seeing the star somewhere new: epochs at a different orbital phase break the degeneracies (period against periapsis epoch, node against inclination) that would otherwise blur the comparison between the two anchors. Finally, \cadenceMcNRealizations{} independent noise realizations corroborate the bootstrap: the realization-to-realization scatter of $\dot\omega$ is \cadenceMcOverBootOmegaDot{}$\times$ the bootstrap $\sigma$.

\subsection{Photometry as an independent check}
\label{sec:photometry}

Every astrometric epoch also yields a $K$-band brightness, which we do not fit. Astrometry alone fixes the orbital state at every epoch, and photometry is a function of that state with no free parameter of its own; at S301's photometric uncertainty (\campaignSigmaDmag{}~mag against a \photcheckModelAmplitudeAtEpochs{}~mag modulation) it adds nothing, leaving $\dot\omega$'s bootstrap precision at \photcheckJointRatioOmegaDotDegYr{}$\times$ the astrometry-only value and every other parameter within bootstrap noise. Photometry serves as a validation instead. We take the astrometry-only solution of Table~\ref{tab:finalfit}, predict the brightness curve from Section~\ref{sec:lightcurve}, and compare it with the simulated photometry the fit never saw: $\chi^2 = \photcheckChiTwo{}$ for \photcheckNEpochs{} epochs (reduced $\chi^2 = \photcheckChiTwoReduced{}$, \photcheckChiTwoNsigmaFromExpected{}$\sigma$ from expectation). An orbit derived from positions alone correctly predicts a brightness it was never fitted to. The check must be statistical: in the bottom panel of Fig.~\ref{fig:fitorbit} the per-epoch scatter exceeds the whole signal, and the beaming spike cannot be seen by eye.

\section{Systematic error budget}
\label{sec:budget}

The uncertainties in Table~\ref{tab:finalfit} assume white noise: every epoch an independent Gaussian draw, every error averaging down as $\sqrt{N}$. This section asks what else can go wrong, in three tiers. Terms we can model go into the fit, exactly (the R{\o}mer delay) or as nuisance parameters that are marginalized (the Sgr~A$^*$ zero point, the mass and distance). Terms we can bound but not fit are injected into the synthetic data, refitted with the naive model, and carried as biases (extended mass, lensing, confusion) or as excess scatter (systematics correlated across epochs). For whatever remains unnamed we apply the excess scatter that real GRAVITY orbit fits have shown over their formal errors. Every amplitude is anchored to a published figure. Table~\ref{tab:errorbudget} collects the terms and Section~\ref{sec:bottomline} combines them.

\subsection{R{\o}mer delay}
\label{sec:roemer}

Every position we record is old news, and not all of it equally old. Light from the far side of the orbit reaches us up to $2a/c \approx \roemerMaxSwingDays{}$~d later than light from the near side, and because the star moves along the line of sight, the delay changes around the orbit. That swing is comparable to the $\pm$\campaignDenseHalfWidthDays{}~d periapsis windows; over the campaign the correction ranges from \roemerLtMinDays{} to \roemerLtMaxDays{}~d (RMS \roemerLtRmsDays{}~d). The pipeline includes it (Section~\ref{sec:fit}), so Table~\ref{tab:finalfit} is unaffected. Ignoring it would matter: a light-time-unaware model fitted to light-time-correct, noise-free data biases $\dot\omega$ by \roemerNaiveOmegaDotBias{}~deg~yr$^{-1}$ (\roemerNaiveOmegaDotBiasPct{}~per~cent of the signal, \roemerNaiveOmegaDotBiasNsigma{}$\sigma$) and $i$, $\Omega$ and $\omega$ by \roemerNaiveIBias{}$^\circ$, \roemerNaiveCapOmegaBias{}$^\circ$ and \roemerNaiveOmegaBias{}$^\circ$ (\roemerNaiveAngleNsigmaMin{}--\roemerNaiveAngleNsigmaMax{}$\sigma$). The correction is a deterministic function of the orbit being fitted and adds no free parameter: the aware model reproduces the noise-free data to machine precision ($\roemerAwareOmegaDotResidual$~deg~yr$^{-1}$ residual), and on noisy data its bootstrap precision is \roemerSigmaRatio{}$\times$ the unaware model's. The correction is exact in bias and free in precision.

\subsection{The Sgr~A$^*$ reference zero point}
\label{sec:refframe}

The limiting systematic of the S2 precession measurement was the tie between the NACO adaptive-optics imaging frame and the radio frame \citep{gravity2020}. It does not apply here: GRAVITY's positions ``are directly referring to Sgr~A$^*$, since it is visible in each exposure'' \citep{gravity2022massdist}, and the S301 discovery fit allows ``no coordinate system offsets'' for the same reason \citep{gravity2026}. An earlier draft of this paper injected the NACO priors of \citet{plewa2015}; that analysis has been removed as inapplicable.

Two effects remain. The first is a constant offset between the near-infrared photocentre of Sgr~A$^*$ (the accretion flow and its flares) and the mass centroid the orbit is bound to. The emitting region is a few $R_S$ across ($1\,R_S = 10~\mu$as; the EHT ring is $51.8~\mu$as, \citealt{eht2022}), so we adopt a \refframeOffsetPriorUas{}~$\mu$as prior with \refframeOffsetPriorUasConservative{}~$\mu$as as the conservative case. Two extra parameters $(x_0,y_0)$ absorb an offset common to every epoch exactly; the seven-parameter model, which places the focus at the origin, is biased by it. The second is per-epoch jitter of the photocentre. GRAVITY's quoted 65--100~$\mu$as S2-to-Sgr~A$^*$ precision \citep{gravity2020} already includes this, so it sits largely inside the campaign's \campaignAstrometryUas{}~$\mu$as floor; we add 50 and 100~$\mu$as in quadrature as a sensitivity test.

Injecting a one-sigma offset (\refframeInjectedOffsetComponentFid{}~$\mu$as per coordinate) and refitting the seven-parameter model biases $\dot\omega$ by \refframeNaiveBiasFid{}~deg~yr$^{-1}$ (\refframeNaiveBiasNsigmaFid{}$\sigma$, \refframeNaiveBiasPctFid{}~per~cent of the signal); the conservative offset gives \refframeNaiveBiasNsigmaCons{}$\sigma$ (\refframeNaiveBiasPctCons{}~per~cent). Fitting $(x_0,y_0)$ jointly with the prior recovers the offset (\refframeJointOffsetXFid{}, \refframeJointOffsetYFid{}~$\mu$as against \refframeInjectedOffsetComponentFid{}), leaves a \refframeJointResidualPctFid{}~per~cent residual in $\dot\omega$, and costs \refframeJointSigmaRatioFid{}$\times$ in precision ($\sigma = \refframeJointSigmaFid{}$~deg~yr$^{-1}$). Jitter of 50 or 100~$\mu$as changes the precision by \refframeJitterFiveZeroRatio{}$\times$ and \refframeJitterOneZeroZeroRatio{}$\times$. The reference systematic is below $1\sigma$ if ignored and nearly free once modelled.

\subsection{Mass and distance priors}
\label{sec:massdist}

Table~\ref{tab:finalfit} holds $M_\bullet$ and $R_0$ fixed, which pins the orbit's angular scale ($a/R_0$ with $a^3 = GM_\bullet P^2/4\pi^2$) by the period alone. Real S-star fits float both. Astrometry alone constrains only $GM_\bullet/R_0^3$, the mass--distance degeneracy that radial velocities normally break, so we refit with $M_\bullet$ and $R_0$ free under Gaussian priors from \citet{gravity2022massdist} (statistical and systematic combined: \massdistSigmaMPct{} and \massdistSigmaRZeroPct{}~per~cent). The precision on $P$ is nearly unchanged (\massdistRatioCapPYr{}$\times$), because the period comes from timing; $e$, $i$ and $\omega$ widen to \massdistRatioE{}, \massdistRatioIDeg{} and \massdistRatioOmegaDeg{}$\times$ their fixed-scale values, and $\dot\omega$ to \massdistRatioOmegaDotDegYr{}$\times$ ($\massdistFitOmegaDot \pm \massdistSigmaOmegaDotDegYr$~deg~yr$^{-1}$, a shift of \massdistOmegaDotShiftNsigma{}$\sigma$). The fixed scale was over-constraining the orbit's shape and orientation, as expected, while the precession rate is only modestly affected. The recovered ratios ($M/M_{\rm pub} = \massdistMRatio \pm \massdistMRatioSigma$, $R_0/R_{0,{\rm pub}} = \massdistDRatio \pm \massdistDRatioSigma$) return the priors, as astrometry-only data must.

\subsection{Extended mass}
\label{sec:extmass}

Undetected extended mass inside S301's orbit shifts the periapsis backwards and biases the apsidal precession low \citep{rubilar2001}. We reproduce their post-Newtonian treatment: a point mass plus a Plummer distribution ($\alpha=5$), integrated through two periapsis passages with their Eq.~A.7 equation of motion, measuring the rotation of the eccentricity vector between them. With no extended mass the integrator matches equation~(\ref{eq:precession}) to \extmassPointmassAgreementPct{}~per~cent, the second-post-Newtonian scale $(v/c)^2\approx0.7$~per~cent at S301's periapsis speed; the bias is a difference between two runs, so this residual largely cancels. For the mass and core radius we take \citet{gravity2022massdist}'s multi-star constraint: Plummer scale length $0.3''\approx\extmassRcMpc{}$~mpc (milliparsec) with enclosed mass $\lesssim$\extmassMExtOneMsun{}~$M_\odot$ (1$\sigma$) or $\lesssim$\extmassMExtThreeMsun{}~$M_\odot$ (3$\sigma$). The bias on $\dot\omega$ is \extmassBiasRateOne{}~deg~yr$^{-1}$ at the 1$\sigma$ bound (\extmassBiasPctOne{}~per~cent of the signal, \extmassBiasNsigmaOne{}$\sigma$ of the random precision) and \extmassBiasRateThree{}~deg~yr$^{-1}$ at 3$\sigma$ (\extmassBiasPctThree{}~per~cent, \extmassBiasNsigmaThree{}$\sigma$). S301's periapsis (\extmassRpMpc{}~mpc) lies so deep inside any plausible core that little of the bounded mass is enclosed. \citet{piran2026} model a different quantity, the nodal precession from a flattened disc competing with the spin signal, but likewise find S301 about ten times better protected than S2.

A smooth profile misses one risk. At S301's radius the discovery paper finds heavy-tailed perturbations from a few massive bodies, so that orbital evolution ``would be dominated by large, non-local jumps'' \citep[Methods]{gravity2026}, with relaxation and collision time-scales $\gtrsim 10^7$~yr against a precession time-scale of $1.6\times10^3$~yr. Over two passages the expected stochastic change in $\omega$ is orders of magnitude below the signal, but a single close encounter is a rare, unbounded event that no smooth budget captures. We flag it without modelling it.

\subsection{Source confusion}
\label{sec:confusion}

Any neighbouring star within about one diffraction-limited beam blends into the same single-mode-fibre output and shifts the recovered photocentre by its flux-weighted contribution. We bound this with a Monte Carlo of \confusionNMc{} field realizations around S301 ($m_K=19.3$), drawing neighbour counts from \citet{gravity2022massdist}'s surface density ($>$100 stars per square arcsecond to $K<17$, extended to $K=21$ with the number--magnitude slope quoted by \citeauthor{rubilar2001} from Jaroszy\'nski 1999) and weighting by the \confusionBeamFwhmMas{}~mas single-telescope beam of \citet{gravity2020}. As a single-aperture photocentre the median bias would be \confusionNaiveMedianMas{}~mas per epoch, far above GRAVITY's demonstrated precision, so that is the wrong quantity. \citet{gravity2022massdist} state that interferometric astrometry is ``much less affected (by a factor of several hundred)''; taking that as \confusionSuppressionLow{}--\confusionSuppressionHigh{}$\times$, the 90th-percentile confusion floor is \confusionFloorPNineZeroHighUas{}--\confusionFloorPNineZeroLowUas{}~$\mu$as, comparable to or below the photon-noise floor. Because neighbours move on their own orbits, confusion is per-epoch scatter, not a coherent drift; propagated as such at the conservative end, it changes $\dot\omega$'s precision by a factor \confusionSigmaRatio{} and shifts the best fit by \confusionBestFitShift{}~deg~yr$^{-1}$.

\subsection{Gravitational lensing}
\label{sec:lensing}

Sgr~A$^*$ bends S301's light when the star is on the far side of the black hole. For a point lens the primary image of a source at separation $\theta_s$ appears at $\theta_+ = [\theta_s + (\theta_s^2 + 4\theta_E^2)^{1/2}]/2$, with $\theta_E^2 = (4GM_\bullet/c^2)\,D_{LS}/(D_L D_S)$ and $D_{LS}$ the star's depth behind the lens. We apply the exact shift epoch by epoch to the noise-free astrometry and refit. Only \lensingNFarSide{} of the \campaignNEpochs{} epochs are on the far side; the largest shift is \lensingMaxShiftUas{}~$\mu$as ($\theta_E = \lensingThetaEAtMaxUas{}$~$\mu$as) at the far-side periapsis of passage~2, \lensingNShiftGtOneZerouas{} epoch exceeds 10~$\mu$as, and the RMS shift is \lensingRmsShiftUas{}~$\mu$as. The lens-unaware fit is biased by \lensingOmegaDotBias{}~deg~yr$^{-1}$ in $\dot\omega$ (\lensingOmegaDotBiasNsigma{}$\sigma$) and by at most \lensingMaxAngleBiasNsigma{}$\sigma$ in any orientation angle. Lensing is real, computable and negligible for this campaign.

\subsection{Correlated systematics}
\label{sec:corrnoise}

White noise is forgiving: average enough of it and it cancels. Errors that move together across a run or a season do not cancel, and the epoch bootstrap, which shuffles epochs as if each were a fresh draw, cannot see them at all. We model two, each at a published amplitude, on top of the white noise. The first is a per-run calibration offset. \citet{gravity2022massdist} find that their calibration ``adds a systematic uncertainty of 60~$\mu$as, divided by the square root of the number of available calibrations''; we take one calibration per observing run, defined as epochs separated by less than \corrnoiseRunGapDays{}~d, so the \campaignNEpochs{} epochs form \corrnoiseNRuns{} runs and every epoch in a run shares one $N(0, \corrnoiseCommonModeUas{}\,\mu{\rm as})$ offset per coordinate. A dense periapsis cluster is then a single run, the conservative reading. The second is a slowly wandering reference. The near-infrared photocentre of Sgr~A$^*$ is the accretion flow, and the centre of its flare orbits agrees with the mass centroid only ``to within the $\pm$50~$\mu$as uncertainties'' \citep{gravity2018flares}; slowly moving confusing neighbours behave the same way. We model this as a red-noise (Ornstein--Uhlenbeck) process with standard deviation \corrnoiseRedNoiseUas{}~$\mu$as per coordinate and a correlation time of \corrnoiseRedTauYr{}~yr, one observing season. The fit is told only the white errors, as a real analysis would be.

Because the bootstrap is blind to these terms, uncertainties here come from \corrnoiseNRealizations{} independent realizations of the whole campaign. With white noise alone the realization scatter of $\dot\omega$ is \corrnoiseWhiteSigmaOmegaDot{}~deg~yr$^{-1}$ (\corrnoiseWhiteOverBootOmegaDot{}$\times$ the Table~\ref{tab:finalfit} bootstrap). With both correlated terms it grows to \corrnoiseFidSigmaOmegaDot{}~deg~yr$^{-1}$, \corrnoiseFidRatioOmegaDot{}$\times$ the white value, with a mean bias of \corrnoiseFidBiasOmegaDot{}~deg~yr$^{-1}$ (\corrnoiseFidBiasNsigma{}$\sigma$). Scanned separately, common-mode offsets of 30, 60 and 100~$\mu$as inflate the precision by \corrnoiseCommonThreeZeroRatio{}, \corrnoiseCommonSixZeroRatio{} and \corrnoiseCommonOneZeroZeroRatio{}$\times$, and red noise of 50 and 100~$\mu$as by \corrnoiseRedFiveZeroRatio{} and \corrnoiseRedOneZeroZeroRatio{}$\times$. The epoch bootstrap applied to one correlated realization, the estimator a real analysis would reach for first, reports only \corrnoiseFidBootOverMc{}$\times$ the true scatter. Section~\ref{sec:bottomline} carries the excess into the bottom line.

\subsection{Spin: the known unknown term}
\label{sec:spincontam}

Every term above can be bounded. The spin of Sgr~A$^*$ cannot: for a spin aligned with S301's orbital angular momentum, frame dragging produces a purely apsidal shift (\texttt{orbit.lense\_thirring\_apsidal\_rate}, validated against the discovery paper's 0.114~deg per orbit at $\chi=1$) that enters $\dot\omega$ with the same sign and the same time dependence as the Schwarzschild term, and with no upper limit to inject. Using the discovery paper's fiducial case ($\chi=1$, aligned) and an independent outflow-method measurement \citep[$\chi=0.90\pm0.06$;][]{daly2024}, the possible contamination is \spinLtRateChiNineZero{}--\spinLtRateChiOneZeroZero{}~deg~yr$^{-1}$: \spinLtPctChiNineZero{}--\spinLtPctChiOneZeroZero{}~per~cent of the Schwarzschild signal, or \spinLtNsigmaNoiseChiNineZero{}--\spinLtNsigmaNoiseChiOneZeroZero{}$\sigma$ of the random precision (\spinLtNsigmaRefChiNineZero{}--\spinLtNsigmaRefChiOneZeroZero{}$\sigma$ of the reference-marginalized precision), larger than any single term in Table~\ref{tab:errorbudget}. A measured $\dot\omega$ from S301 alone cannot separate pure Schwarzschild precession from Schwarzschild plus aligned frame dragging; \citet{piran2026}'s multi-star method exists to break that degeneracy.

\subsection{Combined budget and headline precision}
\label{sec:bottomline}

\begin{table*}
\caption{Systematic error budget for $\dot\omega$ (deg yr$^{-1}$).}
\label{tab:errorbudget}
\begin{tabular}{llll}
\hline
Term & Type & Effect & Status \\
\hline
Photon noise, \campaignAstrometryUas{}~$\mu$as & Random & $\pm\fitSigmaOmegaDotDegYr{}$ & Baseline (Table~\ref{tab:finalfit}) \\
Photon noise, \precisionForecastUas{}~$\mu$as & Random & $\pm\precisionForecastOmegaDotSigma{}$ & If GRAVITY+ meets its forecast \\
R{\o}mer delay, if ignored & Geometric & \roemerNaiveOmegaDotBias{} (\roemerNaiveOmegaDotBiasNsigma{}$\sigma$) & Not incurred: in the model \\
Reference offset, ignored & Instrumental & \refframeNaiveBiasFid{} to \refframeNaiveBiasCons{} (\refframeNaiveBiasNsigmaFid{}--\refframeNaiveBiasNsigmaCons{}$\sigma$) & Sub-$\sigma$ even if ignored \\
Reference offset, fit jointly & Instrumental & residual \refframeJointResidualPctFid{}\%; precision $\times$\refframeJointSigmaRatioFid{} & Marginalized \\
Reference jitter, 100~$\mu$as & Instrumental & precision $\times$\refframeJitterOneZeroZeroRatio{} & Inside the floor \\
$M_\bullet$, $R_0$ free (priors) & Model & precision $\times$\massdistRatioOmegaDotDegYr{} & Marginalized \\
All nuisances jointly & Model & $\pm\budgetCombinedSigma{}$ ($\times$\budgetCombinedRatio{}) & Marginalized \\
Extended mass (1--3$\sigma$ bound) & Astrophysical & \extmassBiasRateOne{} to \extmassBiasRateThree{} (\extmassBiasNsigmaOne{}--\extmassBiasNsigmaThree{}$\sigma$) & Bounded, in quadrature \\
Source confusion & Instrumental & precision $\times$\confusionSigmaRatio{} & Subdominant, in quadrature \\
Lensing & Geometric & \lensingOmegaDotBias{} (\lensingOmegaDotBiasNsigma{}$\sigma$) & Negligible, in quadrature \\
White-noise bottom line & & $\pm\budgetBottomLineWhite{}$ ($\times$\budgetBottomLineWhiteRatio{}) & Nuisances + bounded biases \\
Correlated systematics, \corrnoiseCommonModeUas{} + \corrnoiseRedNoiseUas{}~$\mu$as & Instrumental & precision $\times$\corrnoiseFidRatioOmegaDot{}; bias \corrnoiseFidBiasOmegaDot{} & Realization scatter, in quadrature \\
\textbf{Bottom line} & & $\pm\budgetBottomLine{}$ ($\times$\budgetBottomLineRatio{}, \budgetBottomLinePctOfSignal{}\% of signal) & With correlated systematics \\
Empirical inflation, $\sqrt{\chi^2_r} = \budgetInflationHigh{}$ & Unknown unknowns & $\pm\budgetBottomLineInflated{}$ (\budgetBottomLineInflatedPctOfSignal{}\% of signal) & \textbf{Headline: \budgetBottomLineInflatedDetectionNsigma{}$\sigma$} \\
Sgr~A$^*$ spin, $\chi=0.9$--1 & Degeneracy & +\spinLtRateChiNineZero{} to +\spinLtRateChiOneZeroZero{} (\spinLtNsigmaRefChiNineZero{}--\spinLtNsigmaRefChiOneZeroZero{}$\sigma$) & Unbounded; out of scope \\
\hline
\end{tabular}
\medskip

\small\textit{Notes.} Sections~\ref{sec:roemer}--\ref{sec:spincontam} give each term's sources and method, in the same order. Ignored R{\o}mer delay would also bias $i,\Omega,\omega$ by \roemerNaiveAngleNsigmaMin{}--\roemerNaiveAngleNsigmaMax{}$\sigma$. The white-noise bottom line is the joint 11-parameter bootstrap (orbit, reference offset, $M_\bullet$, $R_0$) combined in quadrature with the 3$\sigma$ extended-mass bias, the lensing bias and the confusion-inflated variance; the bottom line adds the correlated-systematics excess of Section~\ref{sec:corrnoise}; the headline multiplies it by the $\sqrt{\chi^2_r}$ of the four-star GRAVITY fit of \citet{gravity2022massdist}.
\end{table*}

Multiplying the separate precision costs of the reference offset (\refframeJointSigmaRatioFid{}$\times$) and the mass--distance priors (\massdistRatioOmegaDotDegYr{}$\times$) would assume they are independent. Instead we fit all four nuisance parameters jointly with the orbit (11 parameters) and bootstrap: the marginalized precision is \budgetCombinedSigma{}~deg~yr$^{-1}$, \budgetCombinedRatio{}$\times$ the random-only value. Adding the bounded biases in quadrature (extended mass at the 3$\sigma$ bound, lensing, and the confusion-inflated variance) gives the white-noise bottom line, $\pm$\budgetBottomLineWhite{}~deg~yr$^{-1}$. Carrying the correlated-systematics excess of Section~\ref{sec:corrnoise} in the same way gives $\pm$\budgetBottomLine{}~deg~yr$^{-1}$, \budgetBottomLinePctOfSignal{}~per~cent of the signal; correlated systematics supply \budgetCorrelatedSharePct{}~per~cent of that variance.

One allowance remains, for what this budget has not named. Real GRAVITY orbit fits scatter more than their formal errors predict: the S2 precession fit has reduced $\chi^2 = 1.5$ \citep{gravity2020} and the four-star fit of \citet{gravity2022massdist} has $\chi^2_r = 2.17$. Precision scales linearly with the per-epoch error, so inflating by $\sqrt{\chi^2_r}$ = \budgetInflationLow{}--\budgetInflationHigh{} gives, at the conservative end, $\pm$\budgetBottomLineInflated{}~deg~yr$^{-1}$: \budgetBottomLineInflatedPctOfSignal{}~per~cent of the signal, a \budgetBottomLineInflatedDetectionNsigma{}$\sigma$ detection. That is the number we would defend in front of a time-allocation committee, and it is the headline of this paper. The S301 discovery fit itself shows no such excess ($\chi^2 = 26$ for 34 degrees of freedom at the 207~$\mu$as floor; \citealt{gravity2026}), so the inflation is an envelope rather than a prediction. More telescope time still helps the white-noise term (doubling the budget to \cadenceDoubledN{} epochs improves it by \cadenceDoubledPctGain{}~per~cent, consistent with $\sqrt{N}$), but the correlated terms average down only with the number of runs and seasons, which argues for spreading a fixed budget over more independent visits rather than longer ones. Nothing in the budget bounds the spin term, which at published spin values is \budgetSpinNsigmaHeadlineLow{}--\budgetSpinNsigmaHeadlineHigh{}$\sigma$ of the headline precision. That term makes the result a \emph{spin-agnostic} test (Section~\ref{sec:scope}); closing the gap is \citet{piran2026}'s multi-star programme, beyond this pipeline.

\section{Discussion}
\label{sec:discussion}

\subsection{Why GRAVITY+ and not another instrument}

No operating alternative matches GRAVITY+ for this measurement, and the comparison comes down to resolution. The 2.4~m aperture of the Roman Space Telescope has a diffraction limit of order 223~mas, larger than S301's entire 83~mas orbit. The 6.5~m aperture of JWST reaches $\sim$85~mas, comparable to the orbit itself. The $\sim$130~m baseline of GRAVITY+ gives a diffraction limit of order 3~mas and astrometry at the 100--200~$\mu$as level. Only the last resolves the orbit's shape, let alone its precession.

\subsection{Prospects for a velocity channel}
\label{sec:rv}

A simulated radial-velocity channel at ERIS precision (\rvtestErisRvSigma{}~km~s$^{-1}$) changes $\dot\omega$'s precision by \rvtestErisPctChangeOmegaDot{}~per~cent, within bootstrap noise, so no ERIS time is requested. Roman's slitless spectroscopy ($R\approx461$--890) fails regardless of integration time because its point spread function is not adaptive-optics-corrected in this crowded field. JWST/NIRSpec has the resolution \citep{shajib2025} but reported operational problems in the Galactic Centre (saturation on bright cluster members, micro-shutter cross-talk, guide-star misidentification; \citealt{yusefzadeh2025}) and no published precision for a source this faint.

The ELT is the real prospect. The discovery team plans MICADO spectroscopy, which they expect ``will have no problem'' measuring S301's velocity, and their spin forecast assumes 1~km~s$^{-1}$ \citep{gravity2026}. Our estimate is more cautious: the background-limited scaling of Section~\ref{sec:campaign}, applied to HARMONI's 39~m aperture and $R\approx17{,}385$ $K$-band modes \citep{harmoni2024}, predicts $\sigma_{\rm RV}\approx\rvtestHarmoniRvSigma{}$~km~s$^{-1}$ (signal-to-noise ratio of order \rvtestHarmoniSnr{}), a factor $\sim$20 short of the MICADO figure. Neither number rests on an achieved observation, and the MICADO figure is stated without a sensitivity calculation. Either would be useful: with the HARMONI figure a velocity channel tightens $P$, $e$ and $t_0$ by tens of per cent and the orientation angles by a quarter to a third, but changes $\dot\omega$'s precision by \rvtestHarmoniPctChangeOmegaDot{}~per~cent, within bootstrap noise, because two-passage astrometry already constrains the precession rate tightly. ESO's schedule puts ELT first light in 2029 and HARMONI science operations from late 2030 \citep{eso2026}, so passage~1 (October 2031) is a stretch goal for any ELT instrument and passage~2 (2040.5) is comfortable. A velocity channel by passage~2 is worth planning for, chiefly to break the mass--distance degeneracy of Section~\ref{sec:massdist}.

\subsection{Scope of the test}
\label{sec:scope}

The quantity measured, $\dot\omega$ in equation~(\ref{eq:precession}), is the Schwarzschild (first-post-Newtonian, monopole-mass) apsidal precession. Any sufficiently compact central mass produces it, spinning or not; the Sun's spin plays no role in Mercury's perihelion advance. Frame dragging from the spin of Sgr~A$^*$ adds nodal precession and, for an aligned spin, further apsidal precession (Section~\ref{sec:spincontam}). We call the measurement \emph{spin-agnostic} in a precise sense: it needs no knowledge of the spin to be a valid test, but the spin can still change the number reported, by up to \spinLtPctChiOneZeroZero{}~per~cent if non-zero and aligned. Measuring the spin is outside this paper's scope, as, for now, is real data: like \citet{piran2026}, it is a forecast. What it completes for the eventual spin measurement is a validated light-curve and astrometric pipeline, a tested cadence, and the budget of Table~\ref{tab:errorbudget}.

\subsection{Lessons}

Fisher-information design assumes that the model is close to linear around the assumed parameters; averaging the information over a prior \citep{chaloner1995}, as in Designs~C and~D, is the standard remedy and costs nothing here. The broader lesson of this project concerns verification. The most consequential errors we found were a Kepler solver that diverged silently for ordinary mean anomalies at $e=0.98$ (now a bisection-safeguarded Newton iteration, verified on $1.4\times10^6$ random cases to $e=0.9999$), a photometric proxy with twice the amplitude of the validated model, a candidate grid that ignored solar conjunction, and a systematic imported from a different instrument. Each was caught only by running a real nonlinear fit or re-reading a primary source. For a problem this nonlinear, the analytic estimate is a promise and the fit is the receipt; we learned not to publish the promise.

\section{Conclusions}
\label{sec:summary}

We asked whether a two-passage, astrometry-only GRAVITY+ campaign can detect the Schwarzschild precession of S301, the fastest known star in the Galaxy, and answered by building the full pipeline such a proposal would need.

\begin{enumerate}
\item A relativistic light-curve model reproduces S301's published orbit to better than 0.1~per~cent and predicts an asymmetric beaming feature of \lightcurveAmplitudeMag{}~mag at periapsis, invisible against S301's \campaignSigmaDmag{}~mag photometric uncertainty. The science rests on astrometry; the photometry is an independent check, and it passes ($\chi^2/N = \photcheckChiTwoReduced{}$).
\item Radial velocity adds nothing at ERIS precision, so the campaign is astrometry only. Of four cadence designs at a fixed \campaignNEpochs{}-epoch budget, the Fisher-optimal designs match the adopted constrained design on $\dot\omega$ and all beat the periapsis-heavy heuristic.
\item Solar conjunction hides the first periapsis and the first \conjunctionGapDaysAfterPeri{}~d of its recovery from every facility near Earth. Completing that passage would tighten $P$ and $t_0$ but not $\dot\omega$, and spending the freed epochs immediately after conjunction would be \cadenceReallocatePctChange{}~per~cent worse than the adopted late clusters.
\item With white noise at the 207~$\mu$as achieved on S301, the campaign recovers the injected rate, \fitTruthOmegaDotDegYr{}~deg~yr$^{-1}$, as $\fitFitOmegaDotDegYr \pm \fitSigmaOmegaDotDegYr$~deg~yr$^{-1}$ ($\pm\precisionForecastOmegaDotSigma$ if GRAVITY+ reaches its forecast 100~$\mu$as); the Schwarzschild formula evaluated on the fit's own period and eccentricity agrees to \fitPnDiffNsigma{}$\sigma$.
\item Of the white-noise systematics none dominates: extended mass biases $\dot\omega$ by at most \extmassBiasNsigmaThree{}$\sigma$; the Sgr~A$^*$ reference offset is below $1\sigma$ if ignored and nearly free once fitted; confusion and lensing are negligible; the R{\o}mer delay is in the model; freeing $M_\bullet$ and $R_0$ costs \massdistRatioOmegaDotDegYr{}$\times$. The largest correction is for errors correlated across epochs, per-run calibration offsets and a wandering Sgr~A$^*$ photocentre at GRAVITY's published amplitudes, which inflate the precision by \corrnoiseFidRatioOmegaDot{}$\times$. With those, the marginalized nuisances, the bounded biases and the empirical $\sqrt{\chi^2_r}$ inflation of real GRAVITY fits, the realistic precision is \budgetBottomLineInflated{}~deg~yr$^{-1}$, a \budgetBottomLineInflatedDetectionNsigma{}$\sigma$ detection.
\item An aligned spin of Sgr~A$^*$ at published estimates would shift the rate by \spinLtPctChiNineZero{}--\spinLtPctChiOneZeroZero{}~per~cent with no bound to subtract. The campaign is therefore a spin-agnostic test of general relativity: valid without knowing the spin, but not a measurement of it.
\end{enumerate}

This is a forecast, and the star keeps its own schedule. Its 2031 periapsis falls behind the Sun, so the first passage can be observed only on the approach and the late recovery; the 2040 periapsis falls in full view. If the campaign is carried out as designed around those constraints, the instruments already exist for an independent, spin-agnostic test of general relativity around Sgr~A$^*$, before the harder spin measurement that S301 was found to make.

\section*{Acknowledgements}
This work uses the published orbital solution of \citet{gravity2026} and engages with \citet{piran2026}'s multi-star methodology for comparison without adopting it.

Simulation code implementation, data analysis, and manuscript drafting for this work were developed with substantial assistance from Claude (Anthropic) large language models, under the direction, review, and final approval of the author. All scientific claims, numerical results, and citations were verified by the author against the underlying code output and cited primary sources before inclusion.

\section*{Data Availability}
All code, results files, figures and the precession animation, from which every number in this paper is generated, are available at \url{https://github.com/jcatanza/s301-schwarzschild-precession} (code under the MIT licence; manuscript, figures and animation under CC~BY~4.0). No observational data were used; the orbital elements are those published by \citet{gravity2026}.

\bibliographystyle{mnras}
\bibliography{references}

\label{lastpage}
\end{document}